\documentclass[jgrga]{AGUTeX}

  \usepackage[dvips]{graphicx}
\authorrunninghead{V\"{O}R\"{O}S ET AL.}

\titlerunninghead{Windsock conditioned ram pressure effect}

\begin{document}

%
%

\title{Windsock memory COnditioned RAM (CO-RAM) pressure effect: forced reconnection in the Earth's magnetotail}
%

%
%

\smallskip
\authors{Z. V\"or\"os, \altaffilmark{1,2}
G. Facsk\'{o}, \altaffilmark{3,6}
M. Khodachenko, \altaffilmark{1,4}
I. Honkonen, \altaffilmark{5,6}
P. Janhunen, \altaffilmark{6}
M. Palmroth, \altaffilmark{6}
 }

 \altaffiltext{1}{Space Research Institute, Austrian Academy of Sciences, Graz, Austria.}
 \altaffiltext{2}{Institute for Astro- and Particle Physics, University of Innsbruck, Innsbruck, Austria.}
 \altaffiltext{3}{Geodetic and Geophysical Institute, Research Centre for Astronomy and Earth Sciences, Hungarian Academy of Sciences , Sopron, Hungary.}
  \altaffiltext{4}{Skobeltsyn Institute of Nuclear Physics, Moscow State University, Moscow, Russia.}
  \altaffiltext{5}{NASA GSFC, USA.}
  \altaffiltext{6}{Finnish Meteorological Institute, Helsinki, Finland.}

%
%
%
%


\begin{abstract}

Magnetic reconnection (MR) is a key physical concept explaining the addition of magnetic flux to the  magnetotail and closed flux lines back-motion to the dayside magnetosphere. This scenario elaborated by \citet{dung63}, can explain many aspects of solar wind-magnetosphere interaction processes, including substorms. However, neither the Dungey model nor its numerous modifications were able to explain fully the onset conditions for MR in the tail. In this paper, we introduce new onset conditions for forced MR in the tail.  We call our scenario the "windsock memory conditioned ram pressure effect". Our non-flux-transfer associated forcing is introduced by a combination of large-scale windsock motions exhibiting memory effects and solar wind dynamic pressure actions on the nightside magnetopause during northward oriented IMF. Using global MHD GUMICS-4 simulation results, upstream  data from WIND,  magnetosheath data from Cluster-1 and distant-tail data from the two-probe ARTEMIS mission, we show that the simultaneous occurrence of vertical windsock motions of the magnetotail and enhanced solar wind dynamic pressure introduces strong nightside disturbances, including enhanced electric fields and persistent vertical cross-tail shear flows. These perturbations, associated with a stream interaction region in the solar wind, drive MR in the tail during episodes of northward oriented interplanetary magnetic field (IMF). We detect MR indirectly, observing plasmoids in the tail and ground based signatures of Earthward moving fast flows. We also consider the application to solar system planets and close-in exoplanets, where the proposed scenario can elucidate some new aspects of solar/stellar wind - magnetosphere interactions.
\end{abstract}
%
%

%

\begin{article}

%
%

\section{Introduction}
Magnetic reconnection (MR) is a concept which involves multiple physical processes and can explain fast and energetic explosions in laboratory, space and astrophysical plasmas.
Primarily in space and astrophysical settings huge ranges of spatial and temporal scales are involved which prevent us from understanding MR associated processes in their full multi-scale complexity.
Modeling and experimental efforts usually treat a limited range of scales separately. For example, ion- and electron-scale physics within thin current sheets is needed to understand fast MR in collisionless plasmas. The differential motion of electrons and ions during MR onsets is a clear non-MHD process \citep{sitn13}. On the other hand, the large-scale system-wide reorganization
of field structures, leading to thin current sheets and subsequent MR, is not \citep{sem92}, or only partially considered in MR models, e.g. through boundary (external) conditions \citep{schi93}. The specific boundary conditions which are associated with MR characterize the actual physical system under consideration and its interaction with the plasma environment. From case to case significant differences can exist. Focusing only on planetary environments, we recall that, MR can occur in induced magnetospheres of planets with negligible intrinsic magnetic field (Venus and Mars)  \citep{volw09, east08, zhan12} or in (exo-)planetary magnetospheres with differing magnetic moments at different radial distances (e.g. Mercury,  Earth, Jupiter, Saturn, hot Jupiters) from the Sun (or a star) \citep{russ08, slav10, khod12, anto13}. Therefore, the knowledge of specific external or boundary conditions affecting MR in a given system is very important.
In this paper we study MR in the Earth's magnetotail as forced or triggered by strong boundary perturbations.
\subsection{Forced MR in Dungey substorm model}
The most widely accepted scenario explaining current sheet thinning and MR in the magnetotail is associated with dayside-nightside flux transfer.
In this scenario
MR at the dayside magnetopause occurs preferentially for southward oriented interplanetary magnetic field (IMF). The reconnected flux tubes are carried by the solar wind tailward over the poles. It is a widely accepted view that addition of the magnetic flux to the magnetotail can be convected back towards the dayside magnetosphere due to MR in the tail, or the flux can be stored, increasing the magnetic energy density of the lobes [e.g. \textit{Dungey}, 1963, \textit{Baumjohann and Treumann}, 1997]. As a consequence of loading, the amount of the transported flux increases, leading to strongly enhanced current density in a thin current sheet \citep{schi07}. Thin current sheets with a thickness comparable to the ion inertia length are unstable against internal plasma instabilities, for example MR or current disruption. Multi-point in-situ measurements provide strong experimental evidence that, processes such as the occurrence of southward IMF, flux transport to the tail, formation of a thin current sheet and explosive release of the stored magnetic energy due to unstable current sheet, form a physical sequence of events which can lead to enhanced levels of magnetosphere-ionosphere coupling and substorms \citep{ange08, lui11, naka06}. However, due to the possible localized multiple occurrences and ongoing substorm associated activations, the sequences of such events do not necessarily form a simple causal chain of processes \citep{lin09}. Obviously, simple causalities fail when combinations of external forcing factors act simultaneously over their relevant time-scales, or the internal magnetospheric processes exhibit inertia or memory effects via time-delayed multi-scale coupled responses.
For example, recent statistical results indicate that intervals of high solar wind speed V play a role as a strong secondary driver of daily substorm number \citep{new13}. A possible explanation is that the high-speed solar wind changing over time scales much longer than 1 h drives high temperature \citep{bor98} and high plasma $\beta$ plasma sheet, supporting stronger field-aligned currents and more substorms \citep{new13}. In this way, the daily number of shorter duration substorms can be partially controlled by the slowly-changing global plasma environment in the plasma sheet. Other conditions which can trigger MR or substorm onset include the large-scale magnetic field geometry in the tail, when a "bent tail" configuration or changes in the curvature of magnetic field can lead to the formation of a near-Earth neutral line \citep{kive90}. Both the solar wind speed modulation of the daily substorm number and the bent tail geometry are independent of the polarity of IMF. Therefore, these scenarios are independent of dayside-nightside flux-transfer.
\subsection{COnditioned RAM (CO-RAM) pressure scenario for forced MR in the magnetotail}
We argue here that MR onsets can also be explained by external forcing and memory/inertia effects different from the above described  flux-transfer scenario.
The alternative which we consider here is the mechanism for thin current sheet formation via boundary disturbance or deformation by external solar wind forces without significant flux transfer \citep{hahm85, schi07}. In this scenario large-scale magnetopause perturbations push oppositely directed  field lines towards the neutral sheet where the lines are forced to reconnect. We consider here large-scale perturbations introduced by directional changes of the solar wind flow, driving extended windsock motions of the magnetotail.
Our hypothesis is that the large-scale windsock motions can force MR in the tail when the orientation of the axis of the magnetotail is slowly changing.
Observations show that beyond X=$-150 R_E$ ($R_E$ is the Earth radius), the multiple shifts of the magnetotail axis associated with flow directional changes reach magnitudes of $5-15 R_E$ with characteristic duration times greater than 1 hour \citep{sho96}.
When the solar wind flow direction is not changing,
the complement of the
angle between the magnetopause normal and solar wind flow vector (flaring angle) decreases with down-tail distance, i.e.
the magnetopause surface or the magnetotail axis become quasi-parallel to the wind direction. Thus, the distant tail shape and size are modified by the static
(magnetic and thermal) rather than the dynamic (ram) pressure of the solar wind \citep{has00}. On the other hand, when the solar wind flow direction changes significantly in time, the magnetotail
undergoes a slow adaptation windsock motion realigning itself to the new flow \citep{sho96}.
The combination of changes in the solar wind flow direction and of the time-delayed slow windsock motion can increase the down-tail flaring angle. In this way, the solar wind ram pressure would act not only on the dayside but also on the nightside magnetopause. Obviously, if the magnetotail adapts immediately to a new direction of the solar wind flow, the flaring angle would not change and the ram pressure would not act on the nightside magnetopause.
Since large-scale reorganizations of magnetospheric structures are not immediate, the information about the antecedent external conditions, for example flow directional changes, can be temporarily encoded into the magnetospheric structures. For simplicity, we will refer to this structurally encoded information as "windsock memory". Shortly we introduce the magnetospheric structures which can encode the information about the antecedent external conditions. To this end we recall some global MHD simulation results on SW-magnetosphere interaction processes.

\citet{ser08} used global MHD simulations to study the windsock motions of the tail.
They found two time scales associated with the tail response: fast (10-15 min) and slow (half an hour or longer). The fast tail reaction related to windsock perturbations was explained on the basis of wave/discontinuity propagation effects along the tail. These were interpreted as fast interactions of the propagating discontinuity/wavefront with the tail. In simulations the windsock perturbations are driven by the different SW dynamic pressures acting on northern and southern lobes. The pressure asymmetry is responsible for the vertical windsock shift of the tail. It also drives vertical plasma flows and the $E_Y$ electric field in the tail. \citet{ser08}  noticed that some remnant pressure difference, possibly as a result of current sheet warping, remains until the new equilibrium position of the tail is reached. We think that the interaction of the propagating windsock perturbations with the background plasma, current systems and fields  can lead to large-scale tail deformations or structures (e.g. warped current sheet) which survive the fast passage of the waves. These are the structures which can structurally encode information and explain the slow response of the tail.

Another global MHD simulation by \citet{wal99} describes the response of the magnetotail to the changing orientation of IMF. The authors have also found fast and slow components in the response of the tail. The slow time scale in their simulation is associated with the penetration of the SW electric field ($E_Y$ component) into the magnetosphere seen in the Y-Z plane at X=-20 and -60 $R_E$ \citep{wal99}. The time scale of this response is about an hour. This again indicates that the information about the antecedent SW conditions is retained in slowly evolving structures in the tail.

The structural memory associated with windsock motions can be a key element for understanding the ram pressure associated nightside magnetopause disturbances, which can eventually drive current sheet thinning and force MR in the tail. We call this scenario a 'windsock memory conditioned ram pressure effect'. That is, instead of the usual RAM pressure we are considering here a windsock memory COnditioned-RAM pressure, further referred to as CO-RAM pressure.
In line with the above described global MHD simulation results, several time scales can be involved in this scenario. The 'windsock adaptation/conditioning' time $T_{W}$ is the time scale during which the windsock associated asymmetric dynamic pressure exerted by the SW on the opposite sides of the tail is shifting the tail axis to a new position. The time scale of the flow directional change of the SW is $T_{flow1} \geq T_{W}$.  As the windsock generated perturbation propagates along the tail  wave/discontinuity propagation ($T_{Wave}$) and structure survival (memory) time scales ($T_{Mem}$) can be introduced. $T_{Wave}$ can be defined as the time required for a propagating wavefront to travel a distance or it can be associated with the fast response time scale of the tail. The fast and the slow response times roughly define the adaptation time scale: $T_{W}\sim T_{Mem} + T_{Wave}$. In this paper we consider the case for which $T_{W}\sim hours \geq T_{Mem} \geq 1\; hour > T_{Wave}\sim 10\; minutes$. A subsequent SW directional change with a duration of $T_{flow2}$ can increase the CO-RAM pressure exerted on the nightside magnetopause, assuming that $T_{Mem}>T_{flow2}$ or simply  $T_{flow1}>T_{flow2}$. In other words, when alternating slow and fast SW directional changes are involved the slowly responding magnetotail is unable to follow the faster changes in the solar wind. CO-RAM pressure amplitude and asymmetry depends on the actual solar wind dynamic pressure, involved time scales and on the course and speed of SW directional changes. Some of these parameters can be estimated from multi-point measurements, including the asymmetric pressure driven slowly changing vertical plasma flows or electric fields in the tail. However, we cannot fully reconstruct the large-scale structures which are responsible for memory effects.

CO-RAM pressure asymmetry
leads to oppositely oriented motions of different parts of the tail and the associated strong perturbations can force MR.
We are not aware of any published experimental results that would involve CO-RAM pressure triggered MR in the magnetotail.
However, there exist simulation results which support the idea of non-flux-transfer-associated forced MR in the tail via perturbations or described windsock tail motions associated with MR or memory effects.

Choosing dissipation-free tail boundary perturbations in their analytical and numerical calculations \citet{schi93} demonstrated that near-Earth tail boundary disturbances grow during the tail loading (flux transfer) phase, leading to substorm associated thin current sheets and MR. Magnetopause disturbances with perturbation wavelengths shorter than roughly the magnetopause diameter will generate substorm associated thin current sheets with low probability. For perturbation wavelengths longer than the magnetopause diameter the probability of thin current sheet generation by magnetopause disturbances significantly increases \citep{schi93}.

Recently, it has also been demonstrated by global MHD simulations that sudden flow shears in the solar wind, with typical time scales of tens of seconds, can lead to significant distortions, MR and comet-like disconnections of the magnetotail \citep{boro12}. These short duration flow shears propagate across the magnetosphere separating regions with the "new" and "old" orientations of magnetotail axes without any finite adaptation time or memory effects. Nevertheless, \citet{boro12} noticed that, in his MHD simulations, in addition to the fast shear propagation time scale there is a much longer magnetopause motion time scale, which is associated with flow directional changes in the solar wind. In this paper we are interested in long duration directional changes of solar wind speed and the magnetotail axis.

In order to experimentally determine the sequence of events associated with CO-RAM forced MR, the following steps have to be performed: (1.) search for time intervals of flow directional changes in the solar wind during periods of northward oriented IMF (to exclude the case of flux transfer associated forced MR); (2.) demonstrate how the CO-RAM associated signals propagate across the magnetosphere; (3.) identify the displacements of the tail and determine the  corresponding time-scales related to windsock perturbations; (4.) identify the signatures of forced MR in the tail.
In order to accomplish these goals we will use multi-point (solar wind - magnetosheath - tail) measurements complemented by ground-based data and by global MHD (GUMICS-4) simulations for unique events of tail displacement, as seen repeatedly by two tail probes.

The organization of the paper is as follows. Section 2 describes data, instrumentation and orbits. Section 3 introduces the global MHD model. In Section 4 the solar wind drivers are identified and the windsock motions are described in terms of field, plasma and particle signatures during southward and northward IMF conditions. Section 5 is devoted to the multi-point observations of CO-RAM pressure associated processes and GUMICS-4 simulations of tail motions. Section 6 presents indirect signatures of the CO-RAM pressure forced MR, such as tailward moving plasmoids and Earthward moving bursty flow associated geomagnetic effects. Section 7 summarizes our results and provides a short discussion of  possible CO-RAM pressure effects at other solar system planets and at hot-Jupiters.

\section{Data, instrumentation and orbits}
The time interval between November 18 and November 21, 2010, we selected for further study, offers a good opportunity to analyze magnetotail and auroral region responses to changing direction of the solar wind flow and enhanced ram pressure.

Upstream solar wind plasma and IMF variations were monitored by the WIND spacecraft, post-terminator dusk-side magnetosheath conditions were observed by CLUSTER 1, and the two ARTEMIS probes were in the magnetotail.
The data and instrumentation of different spacecraft is described below.

We use WIND flux-gate magnetometer data (time resolution 60 s) from the  Magnetic Field Investigation (MFI, \citep{lepp95})
and plasma data (time resolution 92 s) from the Solar Wind Experiment (SWE, \citep{ogil95}).
For monitoring dusk-side magnetosheath conditions we use Cluster 1 spin resolution ($\sim$ 4 s) magnetic data from the Flux-Gate Magnetometer (FGM, \citep{balo01}),
and ion data from the Cluster Ion Spectrometry (CIS, \citep{reme01}).
For magnetotail measurements we use data from ARTEMIS probes. The two
ARTEMIS probes formed the part of the THEMIS five spacecraft fleet \citep{ange10}. After finishing
the prime mission goals the THB  and THC probes (further referred to as P1 and P2, respectively)
were gradually placed into stable lunar orbits ( -60 $R_E$ in the deep tail).
We use spin resolution ($\sim$ 3 s) ARTEMIS data from the Flux-Gate Magnetometer (FGM, \citep{aust08}),
and the ion and electron Electrostatic Analyzer (ESA, \citep{mcf08}).
The ESA instruments measure plasma parameters over the energy range from a few
eV up to 25 keV for ions and up to 30 keV for electrons.
Additionally, we use 1 minute time resolution AE-index from the Kyoto WDC and ground-based magnetic observatory data with the same time resolution from Canada (IQA observatory) and Greenland network (SKT, GHB, FHB, NAQ and AMK observatories). The GSM coordinate system is used throughout the paper. GSM stands for Geocentric Solar Magnetospheric system.

Figure 1 shows the trajectories of the three spacecraft (P1, P2 and Cluster 1)  in a rotated coordinate system between 18 November, 00:00 UT and 21 November, 00:00 UT, 2010. The nominal positions of the magnetopause (grey surface) and bow shock (green surface) are also depicted (SSC-4D orbit viewer was used). Over the indicated period of time, the P1 and P2 probes move from the nominal magnetosheath towards the center of the magnetotail and the Cluster-1 probe is in the post-terminator magnetosheath. The probes indicated by points on trajectories correspond to 01:30 UT on November 20, 2010.
\begin{figure}[b]
\noindent\includegraphics[width=20pc]{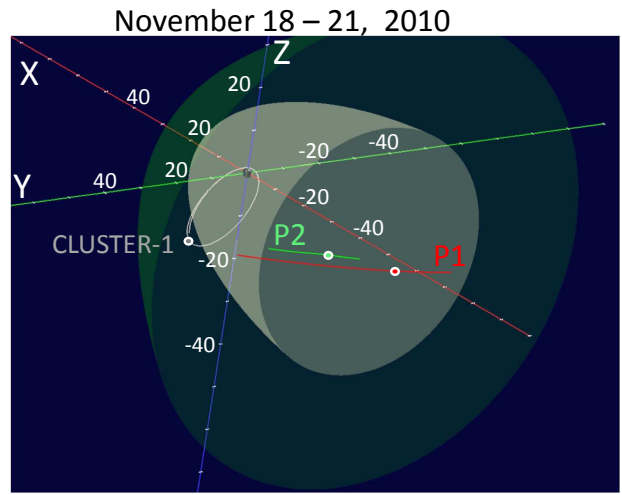}
 \caption{The trajectories of ARTEMIS spacecraft (P1, P2) and Cluster 1 between 18 November, 00:00 UT and 21 November, 00:00 UT, 2010. The nominal positions of the magnetopause and bow shock are  depicted using grey and green colors, respectively (SSC-4D orbit viewer was used). The  points on trajectories correspond to the positions of spacecraft at 01:30 UT on November 20, 2010. These positions are: $X(P1) \sim -67$ $R_E$, $Y(P1) \sim 14.5$ $R_E$, $Z(P1) \sim 3.9$ $R_E$, $X(P2) \sim -53$ $R_E$, $Y(P2) \sim 22$ $R_E$, $Z(P2) \sim 2$ $R_E$ and for CLUSTER-1 $X(C1) \sim -4.4$ $R_E$, $Y(C1) \sim 16.9$ $R_E$, $Z(C1) \sim -11.4$ $R_E$}
 \end{figure}

\section{GUMICS-4 global MHD code}
The GUMICS-4 code represents a global ideal MHD magnetosphere-ionosphere coupling model \citep{janh12}. The simulation box encompasses $\pm$64 $R_E$ in Y-Z directions and extends from X = 32 $R_E$ upstream to X = -224 $R_E$ downstream distances. The solar wind parameters, such as density, temperature, speed and magnetic field are introduced at the upstream boundary and the code reproduces
the magnetic field and plasma parameters inside the simulation box. At 3.7 $R_E$ the magnetospheric simulation is coupled to an electrostatic ionosphere solver through field-aligned currents and electron precipitation. Automatic cell refinement and adaptation is used together with enforced cleaning of magnetic field divergence. More details can be found in the simulation code related publications \citep{janh96, lait06, janh12}.

We mention that in the  global simulation models described in the introduction \citep{schi93, wal99, ser08, boro12} simple stepwise solar wind perturbations were used. The input in our GUMICS-4 simulations is represented by real solar wind data measured by the WIND probe. Our input data are highly variable and this makes the interpretation of global simulation results more difficult.
Nevertheless, we will demonstrate that GUMICS-4 plasma mass density distributions clearly show large-scale windsock associated displacements of the tail axis or smaller scale perturbations of density structures via enhanced ram pressure episodes.

We will also use the simulated conductivity distribution in the auroral region to find the approximate locations of auroral geomagnetic observatories, which observe ground-based signatures of MR associated flows. The capability of GUMICS-4 to simulate plasmoid formation and down-tail propagation has already been demonstrated by \citet{honk11}. However, due to the rather complicated tail dynamics during the windsock events, we will not use GUMICS-4 for plasmoid down-tail motion simulation in this paper.
\section{Solar wind - magnetosphere interactions}
Here we introduce the solar wind conditions and the corresponding magnetospheric responses during crossings of the tail by the P1 and P2 probes. First we consider the total IMF,  the plasma parameters in the solar wind and the measured $V_X$ components of the bulk speed in the tail.
\subsection{Solar wind drivers of the windsock motion}
Figure 2 shows the solar wind conditions observed by WIND and magnetotail bulk speed observations by P1 and P2. The position of WIND is X=256 $R_E$ upstream in the solar wind, at Y = -1 $R_E$ and Z=25 $R_E$. The subplots a-d show WIND data: the total magnetic field ($B$), bulk speed ($V$), proton density ($N$) and temperature ($T$). The area encompassed by blue dashed lines indicates an interaction region between fast and slow solar wind streams (Figure 2b), where the total magnetic field (Figure 2a) and density (Figure 2c) significantly increase. Actually, it is an interaction region between the trailing-edge of the high-speed stream ($\sim$ 350 km/s) with the slower stream leading edge (reaching $\sim$420 km/s). Anyhow, we call it a stream interaction region. Figure 2e shows the evolution of magnetic ($P_B$, blue line), thermal ($P_{T}$, black line) and static ($P_{stat}=P_{B}+P_{T}$, red line) pressures. Figure 2f shows the evolution of dynamic pressure ($P_{dyn}$, black line) together with directional changes of solar wind speed vector ($\phi$, red line). The same Y axis is used for both $P_{dyn}$ and $\phi$.
Within the stream interaction region, first both the static and dynamic pressures increase until $\sim$ 03:00 UT, November 20, then both pressures decrease. The pressure maxima $P_{dyn}\sim 2.2$ [nPa] and $P_{stat} \sim 0.05$ [nPa] show that $P_{dyn} >> P_{stat}$. The increased pressures occurring within the stream interaction region are also associated with substantial directional changes of the solar wind speed vector. The angular change $\phi$ represents any deviation from the radial Sun-Earth direction, including both vertical and azimuthal plasma flow directional changes. $\phi$ changes from 2 degrees (at 06:00 UT on November 19) to 7.5 degrees (after 03:00 UT on November 20). We note that the changes of the aberration angle dependent on the magnitude of the speed V solely are not considered here. The reason is that within the stream interaction region highlighted by blue dashed lines, the changes of V are small ($\sim$ 30 km/s). Moreover, the direction of the flow ($\phi$) correlates rather well with the observed tail displacements (seen in $V_X$ by P1 and P2, explained below).
Global MHD simulations have shown that a significant magnetotail motion and response can be expected for only 6 degrees change of the solar wind flow direction due to the windsock mechanism (gradual realignment of the tail axis to the new direction of the solar wind \citep{ser08}). In our case, the flow directional changes are associated with increased static and dynamic pressures, which make the magnetotail response more complex.

Figures 2g, h show
the $V_X$ component of the plasma speed observed by P1 and P2 along the trajectories across the magnetotail depicted in Figure 1. Initially, both probes are in the magnetosheath (detecting anti-sunward flows of $\sim$ -400 km/s).
First, P2 crosses the magnetopause before 18:00 UT on November 18. P1 follows P2 roughly at 18:00 UT. We will not study this magnetopause crossing in detail. We note that, when the probes enter the magnetotail/magnetosphere, the plasma speed starts fluctuating around $V_X\sim 0$ km/s. The probes enter the magnetosphere during the period of fast stream in the solar wind (Figure 2b). From 18:00 UT, November 18 until 14:00 UT, November 19, P1 and P2 observe $V_X \sim 0$ km/s average flow speeds, while $P_{dyn}$ and $P_{stat}$ pressures in the solar wind do not change significantly. Solar wind flow directional changes are about  $\phi$ = 1-2 degrees
(Figure 2e, f). Between 19:00 UT, November 19 and 12:00 UT, November 20, P1 and P2 detect three time periods of systematic deviation from the average $V_X \sim 0$ km/s in the magnetotail, associated with the largest directional
changes of solar wind flow direction occurring within the stream interaction region (Figure 2 f-h). In what follows, we will refer to these time periods as windsock events A, B, C, respectively. Typical time scales associated with the windsock mechanism are longer than 1 hour \citep{sho96}. The duration of windsock events A, B, C is roughly 6 hours each, and the whole duration of extended magnetotail motion is 18 hours (Figure 2 g, h).

During the event B, P1 and P2 observed flow speeds between $V_X\sim -200$ and $ -400$ km/s. These values are somewhat smaller but comparable  to the speed in the magnetosheath on November 18 or to the solar wind speed within the stream interaction region. It indicates that the magnetotail moved over the position of the probes and as a consequence, P1 and P2 detected magnetosheath-like flows. Since P2 was closer to the nominal magnetopause (Figure 1), it detected stronger flows than P1.
  \begin{figure}[t]
\noindent\includegraphics[width=20pc]{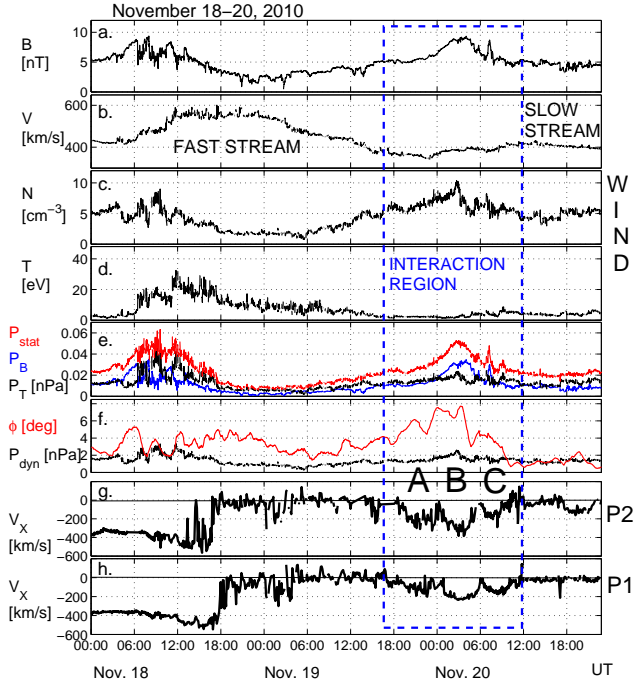}
\caption{The subplots a-f show WIND (solar wind) data: (a.) the total magnetic field ($B$), (b.) bulk speed ($V$), (c.) proton density ($N$), (d.) proton temperature ($T$), (e.) magnetic ($P_B$, blue line), thermal ($P_{T}$, black line) and static ($P_{stat}$, red line) pressures, (f.) dynamic pressure ($P_{dyn}$, black line) and directional changes ($\phi$) of the solar wind speed vector. The subplots g-h show ARTEMIS P1 and P2 tail observations of $V_X$ component of the bulk speed.
The blue dashed box indicates an interaction region between fast and slow solar wind streams. Capital letters A, B and C indicate intervals of windsock motion.
}
 \end{figure}

\begin{figure}[t]
\noindent\includegraphics[width=25pc]{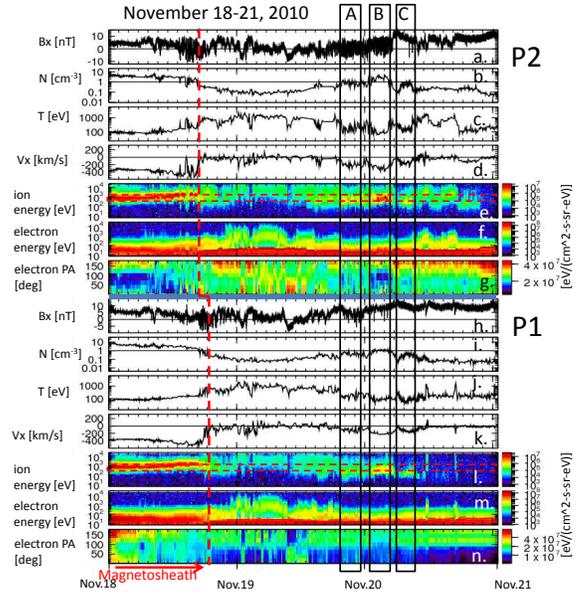}
 \caption{Subplots a-g show P2, subplots h-n P1 data: (a.,h.) $B_X$ component of the magnetic field, (b.,i.) proton density $N$, (c.,j.) proton temperature $T$, (d.,k.) $V_X$ component of the bulk speed, (e.,l.) ion energy distribution, (f.,m.) electron energy distribution, (g.,n.) electron pitch angle distribution. Vertical red dashed lines separate magnetosheath and magnetosphere crossings. Black boxes indicate the occurrence of windsock events A, B and C. }
 \end{figure}

\subsection{Field, plasma and particle signatures of the windsock motion}
In this section we demonstrate that the large-scale windsock motions of the tail can be observed by both P1 and P2 probes during their magnetotail crossings.
The windsock motions move the tail over the probes and, as a consequence, the probes get closer to the magnetosheath. The simplest way to demonstrate this is to show the variation of field, plasma and particle data over the considered tail crossing period, including both the magnetosheath and windsock data intervals. Importantly, the identification of windsock intervals and motion in terms of $V_X$ profile will help us to estimate the windsock adaptation time.

Figure 3 shows magnetic field, plasma and particle observations for P2 (Figure 3 a-g) and P1 (Figure 3 h-n) between November 18 and November 21 2010. The windsock intervals A, B and C are indicated by black vertical boxes. For each probe the $B_X$ component of the magnetic field, ion density (N), ion temperature (T), $V_X$ speed, ion energy spectra and electron energy spectra, and electron pitch angle (PA) distribution are shown. The red vertical dashed lines on November 18 roughly separate magnetosheath/magnetopause crossings during the inbound motion of the probes (left from the vertical line) from nominal magnetotail crossings (right from the vertical line). P1 and P2 show very similar detections: dense and cold magnetosheath, hotter and rarefied magnetotail.
P2 enters the magnetotail earlier than P1. The magnetopause boundary layer represents a transition region between the magnetosheath and magnetosphere, where field and plasma fluctuations take place. Within this transition layer particles can be accelerated. In fact, within the magnetopause transition layer (before the vertical red dashed lines), ion energy spectra (Figures 3 e, l) show that ion energies increase to a few keVs, reaching energies over the typical magnetosheath values. The red dashed horizontal boxes (Figures 3 e, l) roughly indicate the energy range of magnetosheath ions (from $\sim$ hundreds of eV to 1keV).
After entering the magnetotail (vertical red dashed lines) the leading magnetosheath population of ions disappears (within the horizontal red dashed area in Figures 3 e, l) and predominantly the higher energy ions remain within the plasma sheet. Electron energy spectra (Figures 3 f, m) also show higher energy populations in the plasma sheet than in the magnetosheath.
The density between
magnetopause crossings (dashed red vertical lines) and the windsock event A varies between N = 0.1 and 1 $cm^{-3}$. Although these densities indicate plasma sheet or plasma sheet boundary layer values, short encounters with tail lobes are also observed. Tail lobe crossings can easily be identified as intervals of large $abs(B_X)$ with missing high energy ($>$ 2 keV) ion populations \citep{grig12}.

Let us concentrate now on the large-scale windsock events A, B and C in Figure 3.
Only the windsock events show density, ion temperature, speed values (Figures 3 b,c,d,i,j,k) and ion populations within energy ranges (inside the red dashed horizontal boxes in Figures 3 e, l) approaching the values which have been observed in the magnetosheath earlier. Magnetosheath ions appear mainly during event B, when also the strongest anti-sunward magnetosheath like plasma flow is detected. Nevertheless, events A and C show similar correlations. Electron pitch angle distributions (Figures 3 g,n) also confirm that the magnetotail underwent large-scale motions during windsock event observations. In the magnetosheath, both P1 and P2 observed counterstreaming strahl electrons. The strahl is represented by the largest fluxes of electrons near 0 and 180 degrees, indicating electron streams parallel and antiparallel to the magnetic field, respectively. Magnetosheath electron distributions differ significantly from plasma sheet electrons. Counterstreaming strahl electrons on interplanetary magnetic field lines may indicate closed field lines connected with both ends to the Sun \citep{gosl87}, but may also be associated with the bow shock, interplanetary corotating interaction regions, coronal mass ejections and interplanetary shocks (see \citet{ande12} and references therein). More importantly, in our case, the appearance of counterstreaming electrons on magnetotail field lines can indicate open tail field lines connected to the interplanetary magnetic field \citep{oie08}. When the magnetotail is experiencing its windsock motions A, B and C, P1 and P2 repeatedly leaves the closed plasma sheet field lines, observing strahl electrons on open tail field lines. For example, during the windsock event B, the strahl electron population (simultaneous enhanced fluxes of electrons near 0 and 180 degrees) is seen more clearly by P2 than P1, simply because P2 is closer to the magnetopause during the windsock events, therefore it crosses more open field lines. Between the magnetosheath interval and windsock events the peak distribution of pitch angles changes according to the sign of $B_X$, indicating the intermittent visits of northern and southern hemispheres.

Although the field, plasma and particle signatures are slightly different along the trajectories of P1 and P2, we can conclude that the large-scale windsock events A, B and C, seen also as strong tailward flows (- Vx) in Figures 2 g, h and 3 d, k, are observed by both probes. Since the smoothly changing $V_X$  reflect the windsock motion of the tail, we can roughly estimate the windsock adaptation time from the $V_X$ profiles directly.
The $abs(V_X)$ slowly increases then decreases during the windsock events. The durations of these increasing/decreasing phases are between $T_{W}$=0.5 - 2.5 hours.

\subsection{Northward and southward oriented IMF during windsock events}
Let us discuss now how the upstream conditions influence the magnetotail dynamics during predominantly northward IMF ($B_Z > 0$ nT) or southward IMF ($B_Z < 0$ nT) conditions.
Figure 4 shows magnetic field observations upstream in the solar wind (WIND, Figures 4 a,c), in the post terminator magnetosheath (Cluster 1, Figures 4 b,d) and ARTEMIS observations in the magnetotail (Figures 4 e,f,g). Here we show 3-point running mean smoothed ARTEMIS data.
Observations by different spacecraft and the AE-index time series are color coded, which is shown on the top of the Figure 4.
Again, the windsock events are indicated as strong antisunward flows seen in $V_X$, temporarily interrupted by slow flows, depicted together with geomagnetic AE-index in Figure 4h. Figures 4a, b and f show that the negative interplanetary $B_Y$ component penetrates to the magnetosheath and magnetotail.

The strong $B_Y$ components observed in the solar wind and in the tail lead to the following effects: (a.) twisting and flattening of the magnetotail, when the major axis of the deformed elliptically shaped tail tends to be aligned with the IMF; (b.) strong flapping motions of the current sheet, seen as high-frequency sign changing fluctuations in $B_X$ (Figure 4e) during and before windsock events A and B. The $B_Y$ and $B_Z$ magnetic components also show strong correlated fluctuations. Twisting/flattening \citep{sib85} and internally or externally driven flapping motions of the tail \citep{ser03, runo05, sit04, lait07, voro09} are well known from previous studies. The discussion of these complex motions is beyond the scope of this paper and will be studied elsewhere.

The signatures of magnetic flux transfer associated with southward IMF are also recognizable in the time series. Before the end of windsock event B, on November 20 at 05:30 UT, the strong flapping magnetic fluctuations suddenly stop and both P1 and P2 observe  an increasing $B_X$  up to $\sim$ 13 nT (Figures 4e,f,g). At this point the difference between $B_X$ components measured by P1 and P2 suddenly reduces to $\sim$0 nT, indicating the formation of a thicker plasma sheet. The missing high energy ion population after the event B (Figure 3 l) indicates that the probe P1 entered the lobe field.
These large-scale reorganizations of the tail were preceded by a change of the sign of interplanetary $B_Z$ magnetic component (Figure 4c) which reached the magnetosheath with a time delay of $\sim $ 1 h 15 min (Figure 4d).
We note that the correlation between $B_Z$ components observed in the solar wind and magnetosheath is strong. The vertical lines interconnecting the peaks in Figures 4 c and d indicate that the time periods between magnetic structures do not change significantly during the transition from the solar wind to the magnetosheath. However, the amplitude of magnetic fluctuations in magnetosheath is about 3-4 times larger than in the solar wind, indicating strong compression near the magnetopause.
The windsock event C is initially associated with southward $B_Z$, and then with sudden changes of IMF $B_Z$ polarity and enhanced geomagnetic activity (AE index is reaching 190 nT). This chain of events, indicated by the dashed black box in Figure 4, can be interpreted as substorm associated transport of the magnetic flux, mass and energy from dayside toward the nightside magnetosphere, leading to large-scale reconfiguration and temporary inflation of the magnetotail. As a consequence, P1 and P2 stop observing current sheet flapping and their distance from the neutral sheet becomes larger (increasing $B_X$).
\begin{figure}[t]
\noindent\includegraphics[width=20pc]{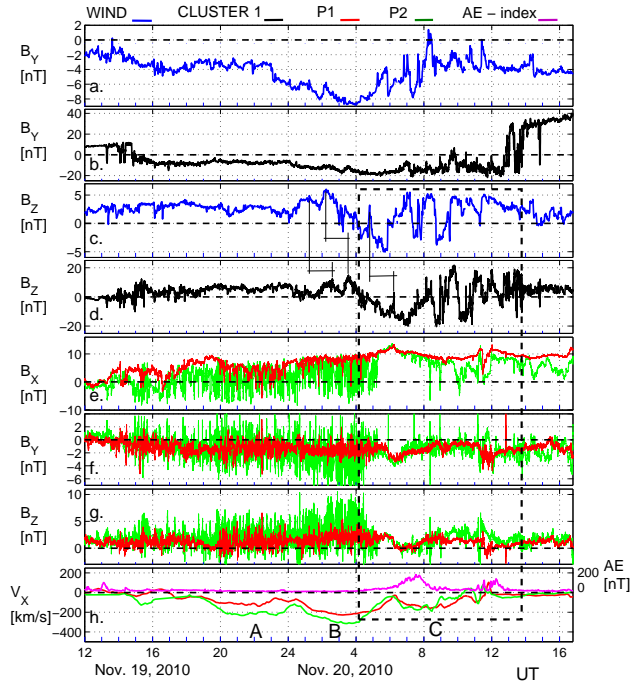}
 \caption{Multi-spacecraft observations of magnetic field (subplots a-g) and bulk speed (subplot h) compared to AE-index (subplot h). (a.,c.) WIND $B_Y$, $B_Z$ (solar wind, blue lines), (b.,d) Cluster-1 $B_Y$, $B_Z$ (magnetosheath, black lines), (e.,f.,g.) P1 and P2 $B_X$, $B_Y$, $B_Z$ (distant tail, red and green lines, respectively), (h.) P1 and P2 $V_X$ and the AE-index (red, green and pink lines, respectively). The color code is indicated on the top of the Figure. The windsock intervals are indicated by capital letters A, B and C. The black dashed box corresponds to the time interval of southward-northward fluctuating IMF. The vertical lines interconnecting peaks in subplots c. and d. indicate correlations between the solar wind and magnetosheath observations.}
 \end{figure}
\begin{figure}[t]
\noindent\includegraphics[width=20pc]{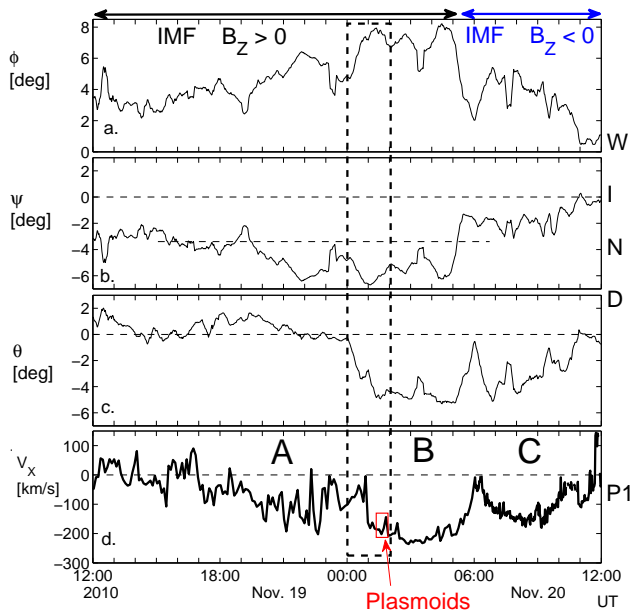}
 \caption{(a.) Directional changes of the solar wind speed vector $\phi$, (b.) azimuthal flow changes $\psi$,  (c.) vertical flow changes $\theta$, (d.) $V_X$ measured by P1 probe in the tail. $\phi$, $\psi$ and $\theta$ are time-shifted to the bow shock. The polarity of IMF is shown on the top. Windsock intervals A, B, and C are indicated by capital letters. Black dashed box shows the interval of largest vertical changes ($\theta$) during northward IMF. Red box shows the interval when plasmoids were observed by P1.}
 \end{figure}

The windsock events A and B occurred predominantly during northward oriented IMF and quiet geomagnetic times (Figures 4 c,d and h, AE index close to zero nT, AU and AL indices as well, not shown).
Figure 4 shows that the IMF is northward between 12:00 UT on November 19, 2010 and 04:00 UT on November 20, 2010, during the windsock event A and partially during the windsock event B.
There exists a short period of negative $B_Z$ at around 00:03 UT on November 20. $B_Z$ reaches -0.8 nT for 10 s and -0.3 nT for 50 s. A statistical study of abrupt IMF turning events (positive/negative IMF $B_Z$s),  with a time duration of at least 45 minutes each, has shown that for $B_Z>-1.5 $ nT the energy input to the magnetosphere is less than 75 GW. For substorm activity input powers exceeding 100 GW are needed \citep{gjer10}. Therefore, the rather short negative $B_Z$ interval is not associated with a flux transport which could lead to current sheet thinning in a situation when $B_Z$ is northward for hours.
Therefore we suppose that during the predominantly northward oriented IMF no significant dayside-nightside flux transfer occurs.

\section{CO-RAM pressure associated perturbations and tail motions}
We demonstrate here that the simultaneously occurring enhancements of solar wind (SW) dynamic pressure and windsock
tail motions can disturb the boundaries of the magnetosphere, shift the magnetotail axis and drive structural reorganizations of the magnetotail. To show this, we use both multi-point measurements and GUMICS-4 simulations.
\subsection{Vertical and azimuthal changes in the direction of the solar wind flow }
Figures 5a-c show the northward and southward IMF associated directional changes of the SW flow
together with windsock events A, B, C, as seen by P1 in $V_X$ observations (Figures 5d). IMF $B_Z$ orientation is indicated on the top of the Figure 5. Here $B_Z<0 nT$ means predominantly southward oriented IMF, including southward-northward polarity changes as well. The flow directional change ($\phi$, Figure 5a) was split into azimuthal ($\psi$, Figure 5b) and vertical ($\theta$, Figure 5c) angles, time-shifted from WIND to the Earth's bow shock. The propagation time from WIND to the bow shock is 1 h 12 min. It can be seen that, during the northward IMF, the events A and B are associated with both azimuthal and vertical flow changes. Event C is associated with  predominantly southward IMF.

The black dashed area in Figure 5 shows a 2 hour long time interval when vertical perturbations and windsock motion can influence tail structures, possibly forcing MR.
Within the black dashed box $\phi$ increased  from 4.3 to 7.5  then decreased to 6.2 degrees. Since vertical motions will bring the oppositely directed field lines from the lobes closer together, more so compared to horizontal motions, changes in vertical flow direction $\theta$ are more important for thin current sheet formation in the tail than changes in horizontal direction $\psi$.  Within the black dashed area  $\theta$ changes from 0 to -5 degrees. These are the largest changes in $\theta$ during the whole northward oriented IMF ($B_Z>0$ nT) interval in Figure 5. As explained earlier there exist large-scale changes in SW flow direction
and windsock motion (Figures 5 a,d) with a duration of $T_{flow1} \sim 6$ h $\geq T_{W}$. Figure 5c shows that $\theta$ within the dashed box changes faster than the slowly changing component, that is  $T_{flow2}(\theta) \sim 2$ h$ < T_{flow1}$.

The area encompassed by red dashed lines in Figure 5d selects a time interval
when magnetic signatures of potential plasmoids were observed by P1 at the beginning of the event B. We explain plasmoid detection in detail later. There were no other plasmoids detected during the interval of northward oriented IMF.
It can be seen from Figure 5 that plasmoid signatures were preceded by the largest vertical flow directional changes in both $\psi$ and $\theta$ within the black dashed box. As was previously explained, such changes
can lead to asymmetric CO-RAM pressure exerted on the opposite lobes.

\subsection{Multi-point measurements: windsock associated pressure, cross-tail flows and electric field}
It was shown by \citet{ser08} that, in global MHD simulations, the adaptation and motion of the magnetotail to the new direction of the SW flow is accompanied
by total pressure differences in the opposite sides of the lobes, buildup of electric fields and significant cross-tail plasma flows. Let us check how the corresponding data available from multi-point measurements change in time and space. To study the relationships between windsock motions and tail response we consider the data within the dashed box in Figure 5.

Figure 6a shows the orientation of the SW flow vector (angle $\phi$, time-shifted from WIND to the Earth's bow shock) and Figures 6 b,c show the $V_Y$ and $V_Z$ components of the bulk speed measured by P1 and P2 in the tail.

The values of the dynamic pressure $P_{dyn}$ computed from WIND and Cluster-1 data are depicted in Figure 6d. Figure 6e shows the YZ components of the electric field
$E_{YZ}= - [\mathbf{V}\times \mathbf{B}]_{YZ}$.  By considering only the windsock associated cross-tail flows the main contribution comes from $V_{Z}B_{X}$ and $V_{Y}B_{X}$ components. Observations by different spacecraft are 3-point running mean smoothed data and are color coded, which is indicated on the top of the Figure 6.
Considering both $V_{Y}$ and $V_{Z}$ components in $E_{YZ}$ allows us to
study the integrated tail response associated with vertical and azimuthal changes of SW flows. Since the magnetotail is tilted,
the GSM $V_{Y}$ and $V_{Z}$ components do not correspond to pure  azimuthal or vertical plasma flows. Unfortunately, due to strong tail flapping motions,
the tilt angle cannot be straightforwardly estimated from the data. Nevertheless, GUMICS-4 simulations indicate (not shown) that the tilt of the plasma sheet relative to Y axis is about 25 degrees. Therefore, we still expect that with regard to forced MR, vertical directional changes of the SW are more important than the azimuthal ones.

The dashed vertical blue line at 00:14:30 UT on November 20  shows the beginning of increasing dynamic pressures (static pressures are negligible) in the SW (blue line) and in the magnetosheath (black line) over the background levels (dashed horizontal lines) in Figure 6 d. This is well correlated with the beginning of the interval of directional changes of the SW flow in Figure 6a. The magnetosheath pressure (Figure 6d) remains enhanced until plasmoid signatures are detected by P1. Plasmoid detection times are indicated by red arrows and dashed vertical red lines. The SW pressure remains at enhanced levels until the end of the considered time interval in Figure 6. The post-terminator pressure perturbance observed by Cluster-1 reaches P1 approximately within 26 minutes and P2 within 48 minutes. These are calculated perturbation propagation times based on local Alfven speeds at the positions of P1 and P2. On November 20, between 00:15 and 01:30, the average Alfv\'{e}n speed derived from P1 measurements is 250 km/s while from P2 it is 105 km/s. The geometry and the positions of the spacecraft in X-Y plane are shown in Figure 7. Remarkably, the $E_{YZ}$ electric fields start to increase when the Alfv\'{e}nic perturbations reach the positions of P1 and P2 (Figure 6e).
Although P1 is at larger distance from Cluster-1 than P2 (Figure 7), the Alfv\'{e}nic travel time is shorter to P1 due to the higher Alfv\'{e}n speeds at the regions adjacent to the plasma sheet \citep{ser08}. In our case, P2 is closer to the magnetotail flank than P1. These observations confirm that the windsock perturbations involve wave propagation effects. According to simulations the wave interactions are associated with the rapid tail response \citet{ser08}.
However, it is easy to notice that simultaneously with the build-up of the electric fields P1 and P2 observe strong persistent cross-tail flows. The electric field (0.2-0.6 mV/m in Figure 6e) is of the same order of magnitude as seen in global simulations of the tail response to SW directional discontinuity by \citet{ser08}. At the position of both tail probes $V_Y$ changes from $\sim 0$  km/s to -40 km/s (Figure 6b).  Both P1 and P2 see a similar $V_Y$ profile, indicating the occurrence of large-scale cross-tail flow structure. During the interval of the enhanced magnetosheath pressure, P2 observes positive and P1 negative $V_Z$ vertical flows (Figure 6c). This strong vertical flow shear (the difference of $V_Z$'s between P1 and P2) becomes stronger when after the passage of wavy perturbations the $E_{YZ}$ electric fields start increasing in Figure 6e.
It can be seen from Figure 6 that the enhanced electric field and the persistent vertical flow shear over the spatial separation of P1 and P2 probes ($\Delta X \sim 15$ $R_E$) does exist over the time scale of $T_{Mem}\sim 2$ hours $\sim T_W$.
Although it is not possible to fully separate the short duration wave interaction effects from longer existing structural reorganizations of the tail, we interpret the persistent electric field and vertical flows as structures which determine the windsock associated memory. These structures are generated by the CO-RAM pressure asymmetry acting on the opposite sides of the tail.
We mention that in global MHD simulations of \citet{boro12}, short duration flow shears move along the magnetotail with the SW speed. Abrupt velocity shears which are frequently observed in the SW would drive flow shears in the tail with typical passage time of $T_{Wave}\sim$ minutes over the separation distance of P1 and P2.
Since $T_{Mem}\gg T_{Wave}$, the long duration flow shear in the tail is associated with slow windsock motion of the tail rather than with an abrupt velocity shear.

An important aspect of boundary perturbations is related to dynamic pressure fluctuations. Between 00:45 and 01:50 UT on November 20, 2010  the pressure fluctuations
represent $\sim$5-20\% of the mean values. The background pressures are  indicated by dashed horizontal lines in Figure 6d. For example, Cluster-1 observes a local peak of dynamic pressure in the magnetosheath at 01:23 UT. This is indicated by the black arrow and vertical dashed black line in Figure 6. Using GUMICS-4 global simulations, we can demonstrate the effects of such small-scale pressure fluctuations on the magnetotail and compare them to the large-scale CO-RAM pressure induced tail motions.
\begin{figure}[t]
\noindent\includegraphics[width=20pc]{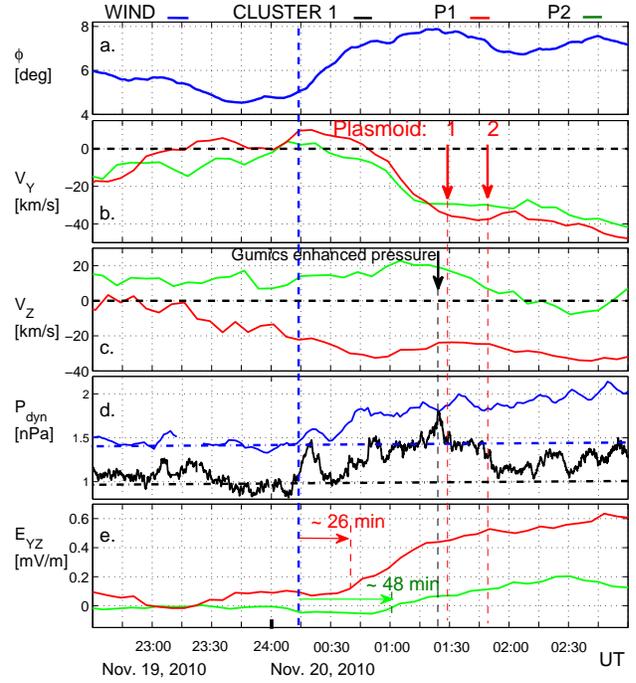}
 \caption{Multi-spacecraft observations: WIND (blue lines, time-shifted to the bowshock), Cluster-1 (black line), P1 (red lines), P2 (green lines). The color code is indicated on the top of the Figure. (a.) directional changes $\phi$, (b.) $V_Y$,(c.) $V_Z$, (d.) dynamic pressure $P_{dyn}$ computed from WIND and Cluster-1, (e.)  electric field $E_{YZ}= - [\mathbf{V}\times \mathbf{B}]_{YZ}$ observed by P1 and P2. The vertical dashed blue line indicates roughly the beginning of the flow directional changes in subplot a., which is well correlated with the beginning of pressure enhancements in subplot d.
 In subplot e. the Alven travel time of the pressure disturbances from Cluster-1 to P1 and P2 probes is shown. The GUMICS-4 enhanced pressure (Figure 7b, see later) associated with dynamic pressure peak observed by Cluster-1 is indicated by vertical black arrow and dashed line. The time instants of plasmoids observed by P1 are indicated by red vertical arrows and dashed lines.}
 \end{figure}
\begin{figure}[t]
\noindent\includegraphics[width=20pc]{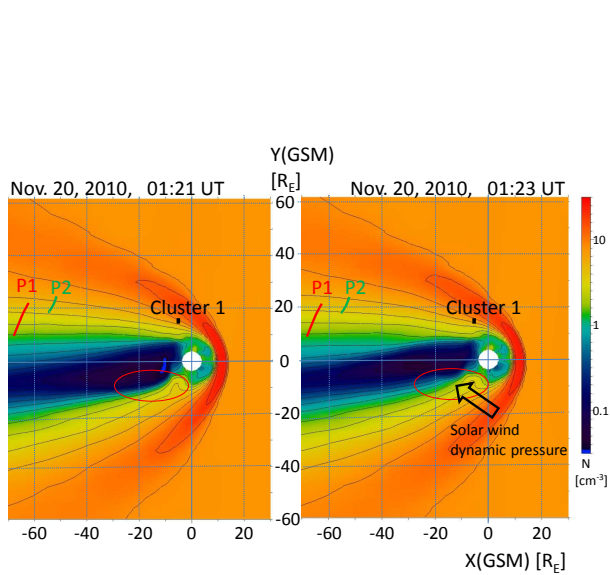}
 \caption{GUMICS-4 density snapshots in X-Y plane.
 Left panel: tail at 01:21 UT on November 20, 2010, before the peak of pressure enhancements. Right panel: the effect of enhanced dynamic pressure on near-Earth tail structures at 01:23 UT. The red ovals show that part of the nightside magnetosphere, which is contracted after the interaction with the propagating solar wind dynamic pressure front. The time when GUMICS-4 shows the tail deformation (01:23 UT) is indicated by vertical black dashed line and arrow in Figure 6. This is also the time when  Cluster-1 in the magnetosheath observes the pressure peak (Figure 6d).}
 \end{figure}

\subsection{GUMICS-4 simulations of tail motions}
First we show the simulation results for the case of small amplitude and short duration pressure fluctuations. We consider the above mentioned pressure pulse observed by Cluster-1 at 01:23 UT.
Figure 7 shows GUMICS-4 results of density distribution in the X-Y plane. The two snapshots, left and right,
correspond to the times 01:21 and 01:23 UT on November 20, respectively. Since the plasma sheet is tilted ($B_Y\neq 0$ nT) and is partially below the equatorial plane, the magnetotail is relatively thin in the X-Y plane.
There is a propagating density front reaching the bow shock /magnetopause approximately at $\sim$ 01:17 UT. The proton density changes by 0.5 $cm^{-3}$ only, therefore it is not seen in the solar wind or magnetosheath in GUMICS-4 snapshots.

The time separation between the two snapshots in Figure 7 is only 2 minutes. The perturbation of
tail density structures shows a spatial size less than 5$R_E$. Taking the Earth's cross-tail magnetopause diameter $D_{ME} \sim$ 30 $R_E$ at X=-10 $R_E$, the perturbation size is less than 0.2$D_{ME}$.
The near-Earth dawn-side magnetotail, indicated by red oval in Figure 7, undergoes contraction
when the dynamic pressure perturbation reaches it at 01:23 UT. Approximately at the same time Cluster-1 observes a small amplitude dynamic pressure enhancement by 0.7 nPa (Figure 6d). The small amplitude pressure perturbation generates a small-scale tail response.  It can be seen from the simulation snapshot that the pressure fluctuation induced spatial perturbations are still rather limited in space and presumably do not lead to thin current sheet generation or forced MR. Anyhow, dynamic pressure fluctuations would not directly affect the nightside tail if it was aligned with the direction of the SW flow. However, small-scale pressure inhomogeneities and fluctuations could raise propagating perturbations along the magnetotail.

We expect that the large-scale perturbations of the tail which are associated with CO-RAM pressure asymmetries are comparable to the tail diameter. As it is explained above in Figures 5 and 6, the directional changes of the SW flow  induce large-scale windsock motions of the tail associated with enhanced dynamic pressure, cross-tail flows, vertical flow shears and build-up of electric fields. Since vertical perturbations of the tail are more important for thin current sheet formation and forced MR than azimuthal perturbations, we consider again the time interval with strongest vertical directional changes of SW flow (dashed box in Figure 5).
Figure 8 shows GUMICS-4 simulation results of density distribution in the X-Z plane. The two snapshots
correspond to the times 00:01 and 01:41 UT on November 20, respectively. It can be seen that the vertical motion and reconfiguration of magnetotail structures associated with changes in the direction of SW flow in Figure 8 are much more significant than the small-scale perturbations driven by pressure fluctuations in Figure 7. As a matter of fact, pressure fluctuations occur continuously
during windsock events A and B and northward oriented IMF. However, potential MR associated plasmoid signatures are detected only when the largest vertical reorientations of the SW flow occur (within the dashed box in Figure 5).

\section{Indirect observation of CO-RAM pressure forced MR}
Although we use upstream, magnetosheath and magnetotail multi-point data in our study, direct observations of MR X-line are not available. An ongoing MR in the tail is recognized indirectly, on the basis of MR associated tailward moving plasmoid signatures and Earthward moving bursty flow driven geomagnetic effects.
\begin{figure}[t]
\noindent\includegraphics[width=20pc]{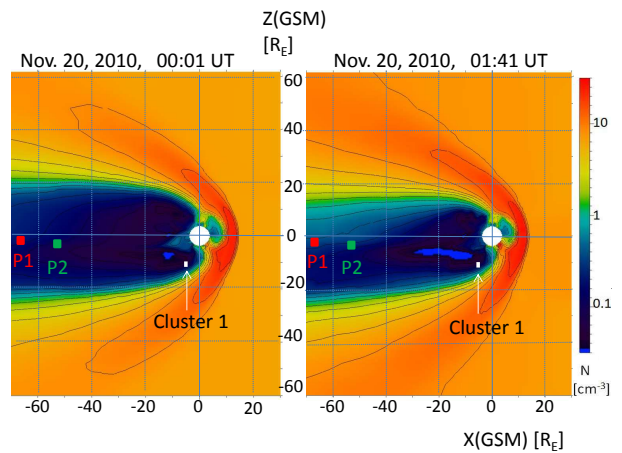}
 \caption{GUMICS-4 density snapshots in X-Z plane. The large-scale reorganizations of the magnetotail are driven by vertical directional changes of the solar wind flow. Left panel: tail before the vertical windsock motions (at the left border of the dashed black box in Figure 5c). Right panel: tail during vertical windsock motions (within the interval indicated by the red box in Figure 5d).
The trajectories of the spacecraft correspond to the time intervals of windsock events A, B and C. }
 \end{figure}
\begin{figure}[t]
\noindent\includegraphics[width=20pc]{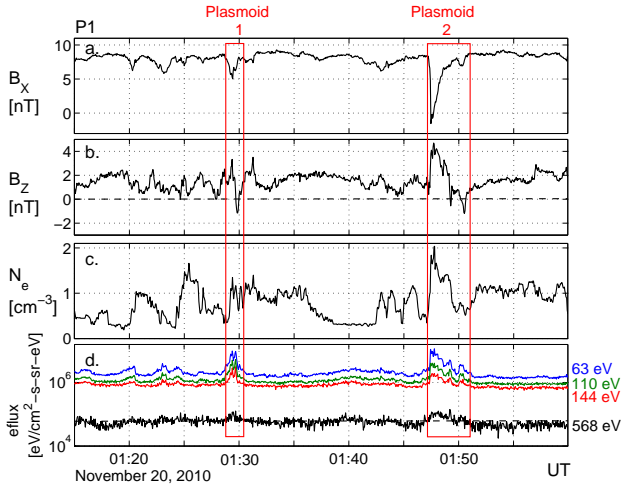}
 \caption{Plasmoid 1 and 2 observations by P1. (a.) $B_X$ component of the magnetic field, (b.) $B_Z$ component of the magnetic field, (c.) electron density $N_e$, (d.) electron flux at different energy channels. Further analysis shows that only plasmoid 2 is associated with ground-based signatures of Earthward moving fast flows.}
 \end{figure}

\subsection{Observation of forced MR associated plasmoids}

The positions of the ARTEMIS probes during plasmoid detections are $X(P1) \sim 67$ $R_E$, $Y(P1) \sim 14.5$ $R_E$, $Z(P1) \sim 3.9$ $R_E$ and $X(P2) \sim 53$ $R_E$, $Y(P2) \sim 22$ $R_E$, $Z(P2) \sim 2$ $R_E$, respectively. In the Y direction the P2 probe is much closer to the magnetosheath than P1, therefore P2 observes flapping current sheet motions propagating towards the flank and their interaction with magnetosheath flow generated fluctuations. The discussion of these complex interactions is beyond the scope of this paper.
Since P2 is completely within the fluctuating flapping associated wavefield, magnetic signatures of plasmoids are seen by P1 only. Nevertheless, P1 is also partially in the wavefield and some care is required to discern the real signatures of plasmoids.
There are two potential plasmoids observed by P1, short duration plasmoid 1 at 01:29 UT and plasmoid 2 at 01:47 UT, both within the red box in Figure 5. The indicated times correspond to the beginnings of plasmoid signatures.

Figure 9 shows the plasmoid observations by P1 in detail. The magnetic components $B_X$ and $B_Z$ are depicted in Figures 9 a and b. Detections of two separated potential plasmoids are indicated by red vertical boxes. Temporarily, we label these events as plasmoids, but we further test the expected plasmoid signatures below. Both plasmoids exhibit decreased  $B_X$ and bipolar $B_Z$ signatures (first positive then negative) within the red boxes. These are typical magnetic signatures of plasmoids (e.g. \citet{zong04}), indicating that the corresponding structures move tailward and the MR site(s) can be situated earthward of P1.

In case of plasmoid 2, the bipolar change is accompanied by a local maximum of $B_X$ component at $\sim $ 01:49 UT within the red box. Before and after the local maximum the values of $B_X$ are smaller than at $\sim $ 01:49 UT. The local increase of $B_X$ component roughly occurs when the $B_Z$ component is changing sign. This is a signature of strong core field and helical flux rope structure inside plasmoid 2, associated with nonzero $B_Y$ \citep{zong04}. The internal flux rope structure
is a consequence of the penetration of IMF $B_Y$ to the plasma sheet as it is demonstrated in Figure 4f. In fact, due to the nonzero $B_Y$ the helical flux rope structure differs from a plasmoid with closed embedded magnetic field. The helical flux rope can be connected to the Earth's ionosphere along the flanks \citep{hugh87} or disconnected from the Earth, but with still recognizable magnetic signatures \citep{birn89}.

Electrons which are accelerated at sites of enhanced electron density within reconnection associated magnetic islands should reach suprathermal energies \citep{chen08}. Figures 9 c and d show electron density and flux enhancements associated with plasmoids 1 and 2. Local electron density enhancements are associated with rising fluxes of thermal electrons, up to $\sim$0.5 keV during the bipolar $B_Z$ signatures. P1 observes enhanced electron fluxes mainly over the energy range of 60 - 600 eV (Figure 9d). However, suprathermal electrons ($>$ 20 keV) were not detected during the passage of plasmoids. A possible explanation for the absence of suprathermal components is that energetic electrons can follow open IMF field lines and rapidly escape from plasmoid regions. The pitch angle distributions in Figure 3n provide some evidence that the probe P1 is in the region containing also open IMF field lines.
Another possibility is that the probe P1 is not close enough to the electron acceleration regions embedded within plasmoids. In fact, P1 observes the plasmoids 1 and 2 only remotely at the beginning of windsock event B, when the tail axis is significantly shifted (Figure 8, right panel).

The magnetic signatures of potential plasmoids could also be masked by flapping motions of the current sheet or their interaction with tailward flows near the magnetosheath.
Both plasmoid and flapping associated magnetic fluctuations can be operative over the time period of minutes. In fact, the bipolar $B_Z$ signature can also occur via flapping \citep{tsur87} and can mimic the magnetic geometry associated with plasmoids or flux ropes. We visually compared magnetic component variations and electron data associated with plasmoids observed by P1 (Figure 9) and with multiple flapping motions observed by P1 and P2.

Crossings ($B_X$ sign change) or approaching (decrease of $B$) the neutral sheet via flapping are also accompanied by increased electron flux (as in Figure 9d). However, simultaneous occurrences of propagating plasmoid magnetic signatures ($B_Z$ sign change), enhanced electron density ($>$ 1 [$cm^{-3}$], and electron flux are  unassociated with the more turbulent flapping motions (not shown). On this basis we can say that the magnetic signatures of plasmoids 1 and 2 are not due to flapping motions.

Another significant difference between flapping motions and plasmoid signatures is recognizable from magnetic hodograms. Figures 10 a-d show the total magnetic field and magnetic components associated with plasmoid 2 in Figure 9. The central parts of the time series are highlighted by thick lines showing a magnetic flux rope core with increased $B, B_X$ and $-B_Y$  embedded into a large-scale plasmoid. Magnetic field hodograms are usually examined in MVA coordinates. Since in our case the tail is very dynamic we plot magnetic hodograms in the coordinate system of all three components. Figure 10e shows the hodogram for plasmoid 2 with magnetic signatures in Figures 10 a-d. The beginning and the end of the event are labeled by START and END. The organization of the plasmoid magnetic field to almost-closed-loops and the central flux rope field to open loop is clearly visible.

Contrarily, flapping associated crossings show no organized structures in magnetic hodograms. Figure 10f shows hodograms of the simultaneous observations of flapping motion by P1 and P2 on November 19, 2010, around 22:04 UT. There is a striking difference between hodograms of more-organized plasmoid structures (Figure 10e) and more chaotic flapping motions (Figure 10f). The hodogram corresponding to plasmoid 1 (not shown) exhibits similar features as plasmoid 2, including also more fluctuations without the developed open loop structure in its center.

We stress here again that the plasmoids 1 and 2 seen by P1 at the beginning of the windsock event B are not associated with increased AE-index or substorms and the IMF was oriented predominantly northward. Therefore, we conclude that
these plasmoids could be ejected by forced MR Earthward from P1, as a result of CO-RAM pressure driven thinning current sheet, without significant flux transfer from the SW to the tail.
\begin{figure}[t]
\noindent\includegraphics[width=20pc]{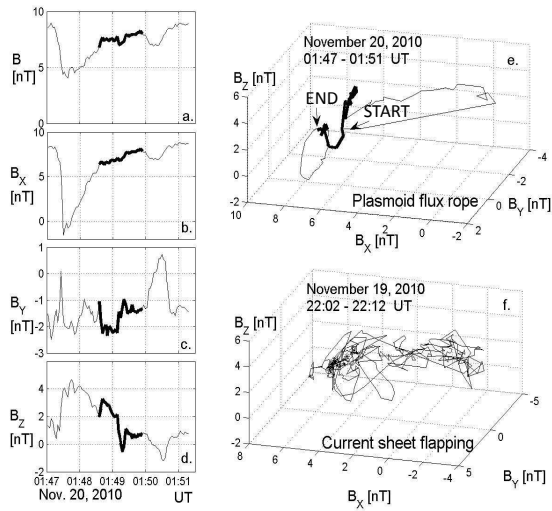}
 \caption{(a.-d. The total magnetic field and its components associated with plasmoid 2 in Figure 8. The thick black line highlights the flux rope structures embedded in the plasmoid. (e.) 3D magnetic hodogram for plasmoid 2.
The labels START and END indicate the beginning and the end of time series, respectively. (f.) 3D magnetic hodogram for a flapping current sheet.}
 \end{figure}
\begin{figure}[t]
\noindent\includegraphics[width=20pc]{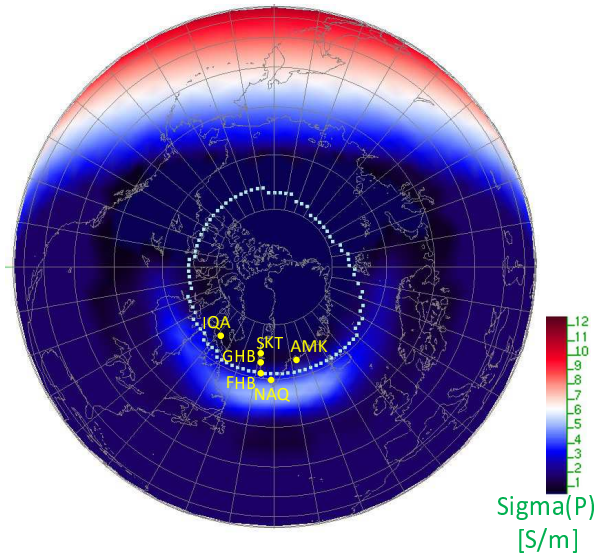}
 \caption{GUMICS-4 simulation of the distribution of Pedersen conductivity over the northern auroral region. The dashed curve shows the border between open and closed field lines. Yellow points are positions of geomagnetic observatories in Greenland and Canada.}
 \end{figure}
\begin{figure}[t]
\noindent\includegraphics[width=20pc]{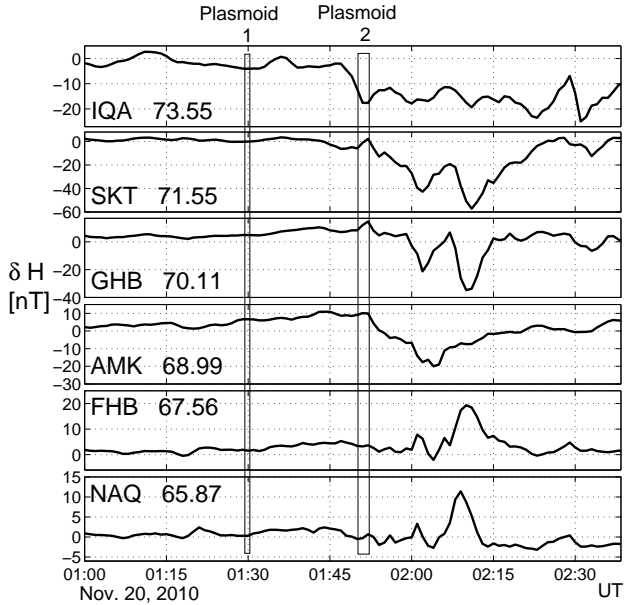}
 \caption{$\delta H$ variations of H magnetic component registered at geomagnetic observatories shown in Figure 11. Vertical boxes show the time of observation of the plasmoids 1 and 2 by P1 probe in the distant tail.}
 \end{figure}
\subsection{Observation of forced MR associated geomagnetic activity}

The forced MR occurring Earthward from P1 can accelerate particles and plasma. Unfortunately, when P1 detected plasmoids, there were no probes in the plasma sheet or at geostationary orbit between P1 and the Earth ionosphere. Therefore, we have no observations of MR accelerated flows or energetic particles.

Earthward oriented reconnection jets or bursty bulk flows can occur during substorm or non-substorm times. In both cases the ionospheric signatures of Earthward propagating bulk flows
are the so-called auroral streamers, which are narrow north-south oriented auroral forms expanding equatorward \citep{amm02}. The reconnection jet and the ionosphere  are interconnected via a field aligned current (FAC) wedge. The FAC system of a reconnection jet can be masked by the strong large-scale substorm current wedge during substorms. In fact, case-studies support the interconnection of the bursty bulk flow current wedge (a small-scale current wedgelet) via FAC with the ionosphere without the development of large-scale substorm current wedge or high ionospheric conductivities \citep{groc04}. However, the plasmoids 1 and 2 detected by P1 are not associated with enhanced AE index or substorm activity. Therefore, we expect that the signatures of non-substorm associated auroral streamers could be detected in ground-based geomagnetic data.

GUMICS-4 simulation of the distribution of Pedersen conductivity and the approximate border between open and closed field line (dashed curve) at 01:50 UT on November 20, are depicted in Figure 11. The simulated Pedersen conductivity is high on the dayside due to the EUV radiation. The nightside non-substorm Pedersen conductivity reaches peak values about 6 S/m centered around midnight over Greenland. The positions of Greenland geomagnetic observatories (SKT, GHB, FHB, NAQ and AMK) and of the Canadian observatory IQA are indicated by yellow points on the map. The ionospheric footpoint of the P1 probe mapped along the field lines is between SKT and IQA stations (not shown). However, precise mapping from the downtail distance of P1 probe (X $\sim$ -67 $R_E$) cannot be expected.

Figure 12 shows the $\delta$H variations of H components (quiet time values were removed) of the magnetic field together with geomagnetic latitudes of the observatories. The times when P1 observed the plasmoids in the distant tail are indicated by vertical boxes.
The northernmost observatory IQA in Figure 12 started to register negative deviations $\delta$ H from the quiet values approximately the same time when P1 started to observe plasmoid 2 at 01:47 UT. Then the negative $\delta$ H disturbance expanded southward, reaching the observatories SKT, GHB and AMK by $\sim 3$ minutes later. At 02:00 UT the southernmost stations FHB and NAQ registered positive  $\delta$ H disturbances. We can speculate that the wedgelet upward-downward FAC current system could develop locally over the stations registering positive and negative disturbances in $\delta$ H, showing also the equatorward expansion features, characteristic for auroral streamers. Another possible ground-based signature of the Earthward moving reconnection jet is the presence of quasi-periodic fluctuations seen in $\delta$ H (less pronounced at AMK). These fluctuations in the range of Pi3 (5-15 min) magnetic pulsations are believed to be the ground-based manifestations of the internal structure of high-speed bursty flows propagating Earthward \citep{solo99}.

It can be seen from Figure 12 that plasmoid 1 is not associated with ground-based geomagnetic effects.
This can indicate that plasmoid 1 or the associated Earthward propagating bursty flows are disconnected from the auroral ionosphere. Alternatively, we cannot exclude the possibility that the short duration plasmoid 1, exhibiting also a more chaotic hodogram than plasmoid 2, corresponds to propagating small-scale CO-RAM pressure fluctuations, as it was suggested in connection with Figure 7. Therefore, it might not be a real plasmoid.

\section{Conclusions}
The main goal of this paper was to show that MR can be forced or triggered in the Earth's magnetotail by external CO-RAM pressure driven disturbances without any significant flux transfer from the dayside magnetopause to the tail. The CO-RAM pressure was defined as a conditioned dynamic pressure acting on the nightside magnetotail when the tail is exhibiting time-delayed windsock motions during episodes of alternating slow and fast directional changes of the SW flow vector. The necessary requirement for CO-RAM is the windsock memory effect, which appears to be the consequence of the temporary survival of magnetospheric structures, including the orientation of the magnetotail, generated by some antecedent SW conditions. Windsock memory is responsible for the temporarily different orientations of the magnetotail axis (or magnetopause) and of the SW bulk flow. This lack of alignment results in an enhanced CO-RAM pressure (even if the normal RAM pressure is unchanged) and large-scale perturbations of tail boundaries, finally leading to the occurrence of forced MR.

In order to test the windsock memory conditioned CO-RAM pressure effect we carried out a multi-point analysis (having probes in the SW, magnetosheath and two probes in the magnetotail) of three long-duration windsock events during extended intervals of directional changes of the SW flow, associated with an interaction region of fast and slow flow streams.  The IMF was initially northward then predominantly southward oriented with nonzero $B_Y$ component penetrating to the magnetotail. IMF $B_Y$ leads to twisted cross-tail field and flapping current sheet. However, these effects will be investigated in a later paper. It was sufficient to show here that the  $B_Y$ driven chaotic flapping motions are rather different from the more organized plasmoid observations in 3D magnetic hodograms. More importantly, the magnetotail underwent significant structural changes at $X \sim -60 R_E$, when the northward oriented IMF changed to fluctuating southward-northward oriented IMF. These changes associated with flux transfer to the tail were followed by moderately enhanced AE-index.

In this paper the main emphasis was put on the interval of northward oriented IMF, when supposedly flux was not transferred to the tail, the AE-index was $\sim$ 0 nT (also the AU and AL indices were $\sim$ 0 nT). The tail was undergoing large-scale windsock motions, however, without those significant structural changes which appeared only after the southward turning of IMF.
Large-scale windsock motions existed under northward IMF conditions for several hours. During this extended period, tailward propagating plasmoid signatures were observed only
when the largest vertical reorientation of SW flow and the corresponding large-scale vertical windsock motion occurred.

CO-RAM pressure associated perturbations (in Z direction) represent a decisive factor in pushing oppositely directed, sunward-antisunward oriented field lines towards the neutral sheet, where the lines are forced to reconnect.
In our case, the vertical windsock motion was associated with enhanced magnetosheath pressure, cross-tail flows, persistent long-duration vertical shear flow structure and enhanced electric fields in the tail. On the basis of the velocity profiles observed by the ARTEMIS probes, which were shown to be generated by the windsock, we estimated the time scale of windsock memory or the windsock
adaptation time to be in the range of $\sim$0.5 - 2.5 hours or less, which is the same or a bit longer time than in global MHD simulations \citep{ser08, wal99}. The windsock associated electric field perturbations were found to propagate faster in the tail center, confirming the results of earlier numerical simulations on the propagation of windsock associated waves \citep{ser08}. The long-duration vertical flow shear in the tail was not seen in global simulations \citep{boro12}, perhaps because of the shortness of the implemented model velocity shears.

One of the observed plasmoids was associated with geomagnetic effects and auroral streamers, these being the ground-based signatures of Earthward propagating bursty bulk flows. Since the tailward moving plasmoid and the ground-based signatures of Earthward moving bursty flows occurred approximately at the same time, these propagating plasma structures were possibly launched by the same MR event in the tail. Actually, this point is not crucial to the issue at hand. Even if at a given time more than one MR sites were launching plasmoids or bursty bulk flows, these MR driven propagating structures occurred during the largest vertical windsock motions and northward oriented IMF. GUMICS-4 simulations have also shown that  large-scale perturbations of the tail during intervals of northward IMF occur when large-scale windsock motions are generated by long-duration vertical changes of SW flow. GUMICS-4 simulations also revealed that short-time low-amplitude dynamic pressure fluctuations can only drive small-scale perturbations of the magnetotail.
In fact, recent CLUSTER and THEMIS spacecraft measurements have indicated that low-amplitude ($\sim 1 R_E$) pressure pulses of different origin in the magnetosheath can perturb the magnetospheric boundary only locally
\citep{ama11, arch12}.

In summary, we interpret the SW flow directional changes, the associated large-scale windsock motions with slow adaptation time and the persistency of cross-tail flow structures and electric field patterns during northward IMF conditions as windsock CO-RAM pressure effect. These processes can include shorter time-scale wave propagation and longer time-scale structurally encoded memory effects. The simultaneous occurrence of CO-RAM pressure generated tail structures, especially the vertical perturbations associated with the tailward-Earthward propagating plasmoid and bursty flows are interpreted in terms of forced MR in the tail, occurring Earthward from the ARTEMIS probes.

Although we are aiming, in particular, to explain the occurrence of forced MR for the case of northward IMF and windsock events, the conclusions of our investigations can be valid for more general cases of varying IMF orientations.
For example, \citet{boud04} have shown that nightside  poleward expansion of the auroral oval can also occur when a dynamic pressure pulse impacts the magnetosphere under conditions of IMF $B_Z \sim 0$ [nT]. In this case the oval expansion is supposedly limited by the amount of stored energy and available magnetic flux in the tail, which can feed and drive MR and magnetospheric convection coupled to aurora. Since $B_Z \sim 0$ [nT] implies no substantial dayside-nightside flux transfer, tail loading could occur earlier when the IMF was southward oriented \citep{boud04}. Similar results on the occurrence of substorms under prolonged northward oriented IMF conditions indicate that
the energy and flux for such substorms could be transferred to the tail during earlier episodes of southward oriented IMF \citep{peng13}. For substorms associated with northward IMF the $B_Y$ component and  the SW dynamic pressure play an important role in the decrease of the AL-index \citep{peng13}. In these investigations \citep{boud04, peng13} the SW dynamic pressure is important. Therefore, we can speculate that our assumption on  forced MR, encompassing the simultaneous windsock motion of the magnetotail with memory and the strong nightside magnetopause disturbance via SW dynamic pressure, is offering a scenario which can explain the release of stored energy in the tail when flux/energy transfer from the dayside magnetopause is absent.
Therefore, there is no need to exclude the case of southward oriented IMF from our considerations. Forcing introduced by the conditioned ram pressure can coexist with dayside-nightside flux transfer (southward IMF) associated forcing, leading to the formation of thin current sheets and forced MR in the magnetotail. The actual orientation of the IMF can modulate the impact of the external disturbances on the tail.
In this respect, further statistical analysis is needed to establish the role of CO-RAM pressure perturbations in magnetotail response or substorms.

Finally, we shortly mention some preliminary ideas about the possible role of CO-RAM pressure induced perturbations in (exo-)planetary magnetospheres. The response of the planets with intrinsic magnetic field to CO-RAM pressure perturbations may depend on several factors such as the magnetic moment of the planets, size of the magnetosphere, internal sources of (co-rotating) plasma (e.g. Jupiter's volcanic moon Io), current systems, radial distance from the Sun (a star), etc. The interplay between these factors can lead to rather complex environments. Nevertheless, with specific regard to the structural memory required for CO-RAM, we can ask the question how long a planetary magnetosphere can store information about the previous states of the SW. Analogously, how fast is the information encoded in structures destroyed, for example, during substorms.
Although the concept of the SW driven Dungey cycle is valid for the giant magnetospheres of Jupiter and Saturn as well \citep{badman07}, the associated time scales are rather long, $\sim$1 month and $\sim$1 week, respectively.
Contrarily, multiple signatures of magnetotail MR, for example at Jupiter, show a much shorter periodicity of 2-3 days \citep{radi10}. It is known that MR in the tail of Jupiter is predominantly internally driven and appears as a result of internal loading and unloading in the rotating Jovian magnetosphere \citep{vas83}. Nevertheless, stereoscopic multi-spacecraft observations have shown that pulses of the SW RAM pressure at Jupiter are strongly correlated with the occurrence of Jovian decametric radio emission \citep{pan13}. It is suggested that RAM pressure pulses at Jupiter are capable to trigger instabilities at the outer edge of the Io torus. The observed bursts of decametric radio emission may originate at regions where interchange instability fingers develop \citep{pan13}.
The unique two-point GALILEO - CASSINI observations during CASSINI flyby of Jupiter on 30 December 2000 have shown that an enhanced SW dynamic pressure front deformed the post-terminator magnetopause of Jupiter at at X$\sim$-25 $R_E$ \citep{kurt02}, causing $\sim$40$R_J$ ($R_J$ is the radius of Jupiter) corrugation of the magnetopause. The dynamic pressure of the SW estimated by \citet{kurt02} reached 0.018 nPa, which is by two orders of magnitude less than in our case at Earth (Figure 6d). Taking the magnetopause diameter of Jupiter $D_{MJ} \sim$ 320 $R_J$ at X=-25 $R_E$, the perturbation size is less than 0.13$D_{MJ}$. These enhanced dynamic pressure generated boundary perturbations resemble the small-scale perturbations at Earth seen in GUMICS-4 simulations (Figure 7). However, there exist no measurements which could support the windsock motion associated CO-RAM pressure scenario for triggering MR at Jupiter. Although there are in-situ and remote observation signatures of ongoing large-scale reorganizations and MR in Jupiter's tail, it is usually not possible to separate the contribution of internal and SW drivers \citep{krupp04}.

External driving can be much more significant for  close-in Jupiter like exoplanets than for solar system giants. The orbital speed of close-in planets is high, therefore the effective speed direction and a magnetotail speed-aligned direction is a vector sum of orbital $V_{ORB}$ and stellar wind speed vectors $V_{STW}$. Hot Jupiters at radial distances 0.05 AU or less from their central stars are known to have circular orbits \citep{gu03}. Therefore $V_{ORB}\sim const.$ and $V_{STW}$ is determined by the structures of source regions (e.g. a coronal hole) on a star. Since  $V_{ORB}$ is high the planet is crossing plasma flow compositions and regions associated with different coronal source areas rapidly. Although a close-in planet encounters relatively hot, dense and not very fast stellar wind, modulations introduced by source regions can lead to significant directional changes of $V_{STW}$ vector, forcing Hot Jupiter magnetospheres via windsock conditioned ram pressure effects. Close-in Hot Jupiter magnetospheres are strongly influenced by stellar radiation and rotation of the planet, leading to expansion and escape of planetary plasma near the equatorial regions. Since the escaping plasma stretches the magnetic field lines, a current-carrying magnetodisk is formed \citep{khod12}. Laboratory experiments and 2D axisymmetric MHD simulations confirm the formation of the magnetodisk and thin current sheet, driven solely by thermal expansion of plasma, even without any centrifugal effects \citep{anto13}. In more realistic situations, a strong external forcing among other windsock memory and tail adaptive motions can influence the reconnection rate at magnetodisk current sheets. This is a rather different physical situation than the interaction of the SW with Jupiter's magnetosphere/magnetodisk in the solar system. Due to the high $V_{ORB}$ the magnetotail of close-in exoplanets can be oriented almost perpendicular to $V_{STW}$ or propagation direction of coronal mass ejections (CMEs). Close-in exoplanets, due to geometric factors, can interact with CMEs much more frequently than the planets in the solar system \citep{khod07}. Global MHD simulations of the interactions of CMEs with close-in planets have shown that the process is more severe and the amount of energy loaded and stored in the magnetosphere is much higher than in the case of the Earth \citep{cohe11}. Yet, with regard to the windsock memory scenario learned from the case of the Earth's magnetosphere, we can argue here that the reaction or adaptation of an exoplanetary magnetosphere would not be immediate. The next forthcoming CME could still encounter the magnetosphere corresponding to the states forced by the former CME.


%
%
%
%
%
%

%
%
%
%

\begin{acknowledgments}
The authors would like to thank A. Runov (UCLA, Los Angeles, USA), R. Nakamura, M. Panchenko, M. Volwerk and T.L. Zhang (IWF, Graz Austria), Ch.J. Owen (MSSL, London, UK), J. Borovsky (University of Michigan, USA) and A. Kendl (University of Innsbruck, Austria) for very useful discussions and support.
This work was supported by the Austrian Fond zur F\"{o}rderung der wissenschaftlichen Forschung (projects P24740-N27,  S11606-N16 and Y398).
Part of the research has received funding from the European Research Council under the European Community's Seventh Framework Programme(FP7/2007-2013)/ERC Starting Grant 200141-QuESpace and from ECLAT FP7 Grant 263325. G\'abor Facsk\'o was also supported by the OTKA Grant K75640 of the Hungarian Scientific Research Fund. We acknowledge NASA contract NAS5-02099 and V. Angelopoulos for use of data from the THEMIS/ARTEMIS mission. We acknowledge the use of Kyoto WDC AE-index and geomagnetic observatory data from IQA (Canada), SKT, GHB, FHB, NAQ and AMK (Greenland). Figure 1 was created by the Satellite Situation Center 4D Orbit Viewer. The GUMICS-4 model developed at FMI represents a part of the computational modeling infrastructure of the EU FP7 project IMPEX (http://impex-FP7.oeaw.ac.at).
\end{acknowledgments}

%
%
%
%
%
%
%
%
%
%




%
%

\end{article}




%
%
%
%
%
%


\end{document}